# Odds Ratios are far from "portable" — A call to use realistic models for effect variation in meta-analysis.

1 June 2021

Mengli Xiao[a], Haitao Chu[a, *], Stephen Cole[b], Yong Chen[c], Richard MacLehose[d], David Richardson[c], Sander Greenland[e]

[a]Division of Biostatistics, School of Public Health, University of Minnesota, Minneapolis, MN 55455, USA

[b]Department of Epidemiology, University of North Carolina, Chapel Hill, NC 27599, USA

[c]Department of Biostatistics, Epidemiology and Informatics, University of Pennsylvania, Philadelphia, PA 19104, USA

[d]Division of Epidemiology and Community Health, School of Public Health, University of Minnesota, Minneapolis, MN 55455, USA

[e]Department of Epidemiology and Department of Statistics, University of California, Los Angeles, 90095, USA

[*]Correspondence: chux0051@umn.edu

**Abstract**

**Objective:** Recently Doi et al. argued that risk ratios should be replaced with odds ratios in clinical research. We disagreed, and empirically documented the lack of portability of odds ratios, while Doi et al. defended their position. In this response we highlight important errors in their position.

**Study Design and Setting:** We counter Doi et al.'s arguments by further examining the correlations of odds ratios, and risk ratios, with baseline risks in 20,198 meta-analyses from the Cochrane Database of Systematic Reviews.

**Results:** Doi et al.'s claim that odds ratios are portable is invalid because 1) their reasoning is circular: they assume a model under which the odds ratio is constant and show that under such a model the odds ratio is portable; 2) the method they advocate to convert odds ratios to risk ratios is biased; 3) their empirical example is readily-refuted by counter-examples of meta-analyses in which the risk ratio is portable but the odds ratio isn't; and 4) they fail to consider the causal determinants of meta-analytic inclusion criteria: Doi et al. mistakenly claim that variation in odds ratios with different baseline risks in meta-analyses is due to collider bias. Empirical comparison between the correlations of odds ratios, and risk ratios, with baseline risks show that the portability of odds ratios and risk ratios varies across settings.

**Conclusion:** The suggestion to replace risk ratios with odds ratios is based on circular reasoning and a confusion of mathematical and empirical results. It is especially misleading for meta-analyses and clinical guidance. Neither the odds ratio nor the risk ratio is universally portable. To address this lack of portability, we reinforce our suggestion to report variation in effect measures *conditioning* on modifying factors such as baseline risk; understanding such variation is essential to patient-centered practice.

## What is New?

### Key findings

- The odds ratio is not portable because it can, and often does, vary with baseline risk.
- The non-collapsibility of the odds ratio can cause disagreements between logistic regression results using different adjustment variables, even when the variables are not confounders. Those variables are commonly unmeasured (e.g., genetic, behavioral, environmental, and clinical factors not measured in a given study) and will render an odds ratio misleading on average.
- The recommendation to replace risk ratios with odds ratios in clinical research should not be followed.

### What this study adds to what was known

- Neither odds ratios nor risk ratios are universally portable.
- The suggestion to replace risk ratios with odds ratios, based on a false claim of portability, is misleading for clinical research, meta-analyses, and clinical practice.
- Risk ratios and risk differences should remain the preferred measures due to their clinical interpretability.

### What is the implication and what should change now?

- To address lack of portability in meta-analyses, one should model the variation in effect measures with baseline risk and study conditions.

**1. Introduction: assuming a model vs. looking at data**

In a previous paper, Doi et al.[1] advised abandoning risk ratios (RRs) in favor of odds ratios (ORs) when reporting study results. We commented by showing how replacing RRs with ORs can be misleading in clinical studies. Doi et al. responded by insisting that RRs lack portability and should be replaced by ORs[2]. They did so by assuming a logistic model without product terms (“interactions”), which forces a constant odds ratio (OR), and then converting this OR to different RRs under varying baseline risks. Their reasoning is circular: they assume the conclusion of their argument rather than present evidence in favor of it. By assuming a model with a constant OR, they have assumed precisely what is under debate: that the OR is always constant (i.e., portable). Based on their *unjustified* portability assumption for the OR, and an unfortunate *misinterpretation* of regression outputs, Doi et al. concluded that RRs are “meaningless” and should be replaced by ORs.

Although we agree that a logistic model is a statistically sensible choice for binary outcomes, we do so not because we think ORs (derived from the estimated model coefficients) are portable, but rather because we recognize that the logistic model is mathematically convenient. It constrains the fitted probability of a binary outcome to a value between 0 and 1 (where it must lie); moreover, it is a model that can be easily fitted using most statistical software packages, it rarely suffers from non-convergence issues, and it can be applied to case-control studies as if the data had been obtained from a cohort study[3].

A logistic model that contains no product term between the exposure of interest and other variables forces the OR for the exposure of interest to stay constant across those other variables, thus making the OR independent of the variation in the baseline risk of disease caused by other variables. Crucially, this independence is a mathematical property forced on the data by the model; it is not property of the actual data or causal mechanisms of action. Doi et al. seem not to realize that the primary objective of scientific inference is to discern patterns in the data, not report patterns forced by convenient statistical models (which are inevitably wrong to some extent). We disagree with Doi et al.'s OR portability assumption because it is only a consequence of an oversimplified statistical model. By adding product terms as needed to capture complicated data patterns[4], a logistic regression model can accommodate variation in ORs with respect to other variables related to baseline disease risks.

When Doi et al.[2] warned users of the "translational failure" of the RR (risk ratio) by computing a RR from a fixed OR at different baseline risks, their premise was that the OR stayed constant under various baseline risks. Following the same logic as Doi et al., if one assumes the RR is constant over a reasonable range of baseline risks, one can also show that ORs computed from a fixed RR vary by baseline risk and could conclude that the OR will have a "translational failure". Doi et al. noted correctly that when baseline risk is 0.6, a RR = 2 is impossible while the OR does not have such a numerical constraint; this only illustrates how there are fewer mathematical possibilities in which RR is constant than in which OR is constant (which is not necessarily a disadvantage of RR)[5,6]. Again, this is solely a mathematical property of the OR, not evidence that the OR is more often constant in reality. But Doi et al. misinterpret this mathematical property as an empirical one and assume that ORs are always portable. That is a mistake: there is empirical evidence in meta-analyses that ORs often vary with baseline risks[6–8]. For example, among 40,243 meta-analyses in the Cochrane Database of Systematic Reviews (henceforth, Cochrane Reviews), we found that the OR lacked portability in most meta-analyses and presented a specific case study[9]. Yet Doi et al.[2] repeatedly assumed a constant OR when converting the OR to RRs under varying baseline risks, ignoring our previous call for empirical evidence supporting their assumption that ORs are portable[9].

Doi et al. concluded that "the stratum-specific RR is meaningless as it depends on what baseline risk changes occur across strata" from re-analyzing the data we presented to illustrate the non-collapsibility of OR. However, they did not acknowledge that those data were generated to *force* the exposure-outcome OR *to be constant* across strata. If the OR is constant over baseline risk, the RR must, mathematically, vary by baseline risk and vice versa. In fact, Doi et

al.'s conclusion easily fails in other examples where the RR is constant across strata. We illustrate this point by fitting Doi et al.'s regression models to the data in Table 1B from our previous paper[9]. The first two data columns of Table 1 show that while the RR of X (and of Z) stayed the same across both the multiple regression and stratified analysis, the values of OR varied. This change of conclusion from Doi et al. is not surprising because Doi et al. only presented an example in which the stratum-specific OR remains constant and stratum-specific RR changes, rather than showing what happens when instead the OR varies and the RR stays constant.

Doi et al. agreed with us that when interpreting effect measures, one should generally condition on baseline risks but persisted in asserting a "RR of 1.04 belongs solely to the baseline risk of 0.96 and therefore lacks any translational value." Doi et al. went further and stated that "the only way to achieve" an effect measure that accounts for baseline risk "is via the use of ORs." This is incorrect both mathematically and empirically. Mathematically, as we stated above, if the OR is constant across baseline risks then RRs will necessarily vary; but if the RR is constant across baseline risk then ORs will necessarily vary. Our earlier example[9] presented a case where both the RR and OR vary by baseline risk, which empirically may be the most common situation. To accommodate this situation, one may estimate effects conditional on baseline risks using a bivariate generalized linear mixed model[10] (BGLMM) and from that one can derive baseline risk-specific estimates of risk differences (RDs), which better serve actual clinical decisions than do RRs or ORs[11,12]. The assumption that ORs are universally portable is thus unnecessary statistically, unsupported empirically, and can misguide clinical research and practice.

**2. OR from logistic regression and its non-collapsibility**

When addressing our comments on the non-collapsibility of ORs, Doi et al. seem to attribute the non-collapsibility to not accounting for other factors in saying that "non-collapsibility is only an issue when one gets the outcome model wrong by refusing to account for easily accounted for outcome heterogeneity." To illustrate their claim, they fitted logistic and log-binomial regression models to the data in Table 1A from our previous paper[9] and treated the stratum-specific variable Z as "an unmeasured independent risk factor". They computed an OR from the logistic model and a RR from the log-binomial model, respectively. After comparing results from multiple-regression models to simple tabular measures stratified by Z, they claimed that only the OR stayed unchanged and concluded that "An OR can therefore be applied to any specific stratum of the population while, on the other hand, the RR is inadmissible." The arguments that they used to explain the non-collapsibility of ORs are incorrect, as we explain below.

**2.1 Misinterpreted effect modification in logistic regression**

Doi et al. argued that "baseline risk and the dependence of the RR on these risks will distort the results" from the log-binomial regression outputs. The basis of their claim was that the RR of Z from the log-binomial regression model with a product term in their Table 3 (RR=3) differed from the RR of Z produced from the simple tabular calculation stratified on X=1 (RR=2). The differences between these two distinct quantities, RR of Z at X=0 (RR=3) and RR of Z at X=1 (RR=2), were Doi et al.'s evidence for inconsistency of RRs. But this difference resulted from their misinterpretation of their statistical model. They fitted a model with a product term and the RR=3 they presented from their model corresponds to the effect of Z in the reference level of X (X=0), not the X=1 stratum. If the product term were correctly incorporated, the RR of Z at X=1 would be the product between the exponentiated product term (i.e., $2/3 \approx 0.67$) and RR=3 (RR of Z at X=0) under the fitted log-binomial model. Then, RR of Z at X=1 would be 2, a value equal to the simple tabular calculation. There is no discrepancy between the RRs from stratified tabular calculation and the RRs estimated from the saturated log-binomial model.

In explaining Table 3, Doi et al. noted that they relied on the non-significant result (P-value > 0.05) from a likelihood-ratio test to justify their decision to omit the product term between X and Z. But the data are purely artificial numbers that were constructed to illustrate the mathematical property of non-collapsibility, not statistical variation. Even if the data were real, large portions of the statistical and scientific community have condemned use of statistical significance, especially to categorize results based on the P-value, compatibility ("confidence") interval[13–15], or other statistical measures[16–18], or to select or exclude terms in a regression model[19]. This problem is worsened in the example because tests of product terms (statistical interactions) are known to have very low statistical power in typical epidemiologic studies[20,21]. Furthermore, excluding a term simply due to a high P-value will bias both marginal and subgroup estimates[4]. Bayesian approaches have been recommended when treatment effect heterogeneity is of interest[22,23].

### 2.2 Biased conversion of ORs to RRs

Doi et al. acknowledged ORs are difficult to interpret and suggested converting ORs to RRs using the method proposed by Zhang and Yu[24]. However, this conversion method typically provides a biased estimate of RR and leads to a poor compatibility interval coverage under common outcomes[25,26], potentially misleading clinical decisions. Such problems can be avoided if one computes a population-averaged (marginal or standardized) measure of association from a logistic or log-binomial regression model[27,28]; a population-averaged RR can be computed from the risks fitted by the population-averaged logistic model[26,28–31], and when the numbers per stratum are small, a population-averaged OR is less susceptible to bias than the covariate conditional model estimates discussed by Doi et al.[32]

### 2.3 Non-collapsibility of the OR is a mathematical property

While Doi et al. argued that the problem of non-collapsibility is "a reflection of an unmeasured independent risk factor"[2], we want to emphasize that this non-collapsibility is a problem of the OR but not of the RR or RD. This shortcoming of the OR is not simply an academic concern because there are always unmeasured risk factors for a disease (e.g., genetic, behavioral, environmental, and clinical factors not measured in a given study) that will render an OR estimate misleading on average without doing the same to a RR or RD estimate. It is thus a defect of the OR that as a marginal summary it introduces non-collapsibility[33,34].

Non-collapsibility arises because the unstratified OR cannot be written as a weighted average of the stratum-specific OR except under special circumstances[35,36] Doi et al. pointed out that the OR of X stayed the same in both multiple regression and stratified analysis as long as the variable Z (which they called "unmeasured independent risk factor") was taken into account. In doing so they were only reiterating that stratum-specific ORs were the same across different strata for that particular example (that was engineered specifically to have a constant OR across strata of Z), without addressing the more general problem of non-collapsibility: that the crude and stratum-specific ORs differed even though Z was not a confounder. The essential measurement of assessing non-collapsibility, crude measures, were not computed by Doi et al.

To obtain a valid answer to whether the effect measures are collapsible in the absence of confounding, one must compare the crude OR and RR (from simple regression or tabular analysis) with the stratum-specific OR and RR (either from a multiple regression or stratified tabular analysis). As shown by the bolded numbers in Table 1, when stratum-specific measures are the same, the corresponding unstratified ORs for X and Z are closer to the null than the stratum-specific OR while crude RRs for X and Z are equal to the stratum-specific RRs. In both datasets, Z and X can correspond to an unmeasured independent risk factor. Doi et al. failed to acknowledge that the non-collapsibility is a unique disadvantage of the OR, and they only attributed the non-collapsibility to not accounting for Z (or X) in the crude effect estimates of X (or Z) when they assert " 'non-collapsibility' is only an issue when one gets the outcome model wrong by refusing to account for easily accounted for outcome heterogeneity." If non-collapsibility only came from "an unmeasured independent risk factor" as Doi et al. implied, then one would expect non-collapsibility of RR in the left side of Table 1; the adjusted or stratified RRs would be different from the crude RRs, but they are not. Therefore, applying regression models only further illustrates the non-collapsibility disadvantage of OR.

While we agree with Doi et al. that "a population OR should not be applied to any specific stratum of the population unless we do not have access to stratum specific ORs," this advice does little to solve problems of non-collapsibility. Again, many risk factors will remain unmeasured in real-world clinical studies, and the coefficients from regression analysis only represent an OR collapsed over all those unmeasured risk factors[37]. If Z is unobserved in Table 1, then the only feasible regression model would be univariate, but that unstratified OR of 2.25 does not correspond to any average of the Z-specific ORs. In contrast the RR from the univariate log-binomial regression is a weighted average of Z-specific RRs[9]. In practical terms, this means that an observed OR can vary beyond the range of ORs that would be seen if the factors were available *even if these factors are not confounders*, whereas the RR cannot. This possibility is not a bias in the observed OR, but we (like many others) regard it as a serious defect in the OR's interpretability[38–40].

Doi et al. stated that "The OR of 2.25 has been interpreted by Chu and colleagues as a bias in the unconditional OR". This is simply another mistake in their paper: We wrote "one could falsely assume there is confounding bias from OR from Table 1A". We thus emphasized that the non-collapsibility of OR may *falsely* suggest to researchers a confounding bias. There is no bias if one only wishes to estimate the marginal OR. However, we again note that the marginal OR is generally not applicable to stratum-specific ORs. As we mentioned before, the statistical product term was ignored by Doi et al. They seemed to misinterpret an overall OR from the regression coefficient as if it were marginal, when it only represents a stratum-specific estimate. Hence, Doi et al.'s interpretation of the OR from their regression model is another confusion between non-collapsibility and confounding bias.

## 3. Lack of portability of ORs in meta-analysis

### 3.1 Portability is only meaningful for studies within the same meta-analysis

Doi et al. refused to acknowledge that ORs lack portability, despite our previous empirical evidence. We found that ORs were not portable in most meta-analyses among 40,243 meta-analyses in Cochrane Reviews[9]. This evidence contradicts the conclusion from Doi et al.'s first paper in which they mixed distinct meta-analyses together such that the dependence of an OR on the baseline risk is confounded. Doi et al. argued that our previous observation was "spurious" because the strong correlation was from "conditioning on the meta-analysis from which a trial emerges means that Chu and colleagues have conditioned on a collider." However, this claim is absurd. Neither the OR nor the baseline risk determines which meta-analysis they appear in – that determination is instead based on the topic of the biomedical investigation. This means that meta-analysis membership is **not** a collider and Doi et al.'s claim is simply another mistake.

In contrast to Doi et al.'s merging of studies without regard to topic, conditioning on the meta-analysis ensures that the correlations examined are the ones relevant to researchers. Numerous studies have recommended accounting for correlations between treatment effect and baseline risks *within the same meta-analysis* in the treatment effect model[8,41–44]. It is the variation in treatment effects among *studies of the same treatment and outcome* that motivated assessing the correlation of treatment effects with baseline risks[8], and also led to recognition that the OR was not always portable in meta-analysis[8,45].

### 3.2 Portability of OR and RR depends on specific scenarios

Doi et al. insisted that "Only the meta-analysis of ORs can deliver the correct meta-analytic estimate of the association of X with Y for the two populations defined by Z." Such a conclusion is only valid in scenarios where the OR is portable or constant across strata of Z. Despite their objection to assessing portability within a particular meta-analysis, Doi et al. conducted a meta-analysis of the same example data we discussed above, with a marginal OR=2.25 and constant stratum-specific ORs=2.67, and treated the strata as independent studies. They showed that a meta-analysis of these data results in a meta-estimate OR=2.67.

Again, if we use another dataset, their conclusion immediately fails. The left side of Table 2 shows that while the meta-analysis summary OR is different from stratum-specific ORs, the summary RR stay the same as the stratum-specific RRs (RR=2). Now, following the same reasoning of Doi et al., one may conclude that the OR was “not solely an effect measure because it reflects also baseline risk and varies with it” if the left side of Table 2 is only presented. Based on only one unique dataset, Doi et al.’s that “RR cannot be used for meta-analysis” is misguided.

Similar problems occurred again when Doi et al. constructed a single meta-analysis from real data to support their strong opinion that “RR cannot be used for meta-analysis.” Samples in their constructed meta-analysis were not independent among studies, though Doi et al. used a meta-analysis method to account for the dependency. Moreover, they used a non-significant result from the likelihood-ratio test, known to have a low power[20,21], to claim no variation in the OR.

**3.3 A counter-example with the bivariate generalized linear mixed model (BGLMM)**

Because of those issues, we extend our discussion to a real-world meta-analysis with 20 studies in Cochrane Reviews. Those 20 studies were independent, in contrast to Doi et al.’ analysis of correlated study participants among 4 groups. The selected meta-analysis compared the effect of Radix Sophorae flavescentis treatment (> 6 months) with placebo or intervention on the existence of hepatitis B e-antigen. We will use BGLMM to obtain estimates conditioning on baseline risks to account for potential correlation with baseline risks.

We first calculated Spearman’s rank correlation coefficient (Spearman’s $\rho$) between three common binary effect measures, i.e., OR, RR, and risk differences (RD), and the baseline risk within the same meta-analysis. Calculations for the compatibility interval for Spearman’s $\rho$ were presented in Xiao et al.[9] We applied PROC NLMIXED from SAS (version 9.4) to implement the BGLMM. To justify the validity of the BGLMM, a two-stage random-effects meta-analysis[46,47] (i.e., compute study-specific estimates and their standard errors first, then followed by a univariate random-effects meta-analysis) was used to compare with the marginal OR, RR and RD from BGLMM. We used the "rma" command from "metafor'' package in R by using restricted maximum likelihood method to estimate the between-study heterogeneity[47]. More details about the BGLMM formulation can be found in Xiao et al.[9]

The BGLMMs fit the data well under all measures, evaluated by the criteria that most observed study effects were within the BGLMM prediction intervals (Figure 1). The BGLMM also provides similar estimates of effect measures to those from the two-stage random-effects model. The marginal estimates for OR, RR, and RD are 0.31 (95% CI, 0.23 to 0.40), 0.70 (95% CI, 0.65 to 0.75) and −0.24 (95% CI, −0.29 to −0.20) under the BGLMMs. Those values are close to the two-stage meta-analysis model estimates of 0.34 (95% CI, 0.28 to 0.43), 0.72 (95% CI, 0.67 to 0.77), and −0.24 (95% CI, −0.29 to −0.19) for OR, RR, and RD.

Figure 1 shows the BGLMM fits of OR, RR and, RD conditioning on baseline risks and the corresponding prediction and compatibility ("confidence") regions. The increasing baseline risk brings a noticeable decrease in the OR (Figure 1(a)) while RR and RD were portable across varying baseline risks (Figure 1(b) and (c)). Such a distinct lack of portability of the OR is also supported by Spearman's $\rho$. Spearman's $\rho$ estimates for the correlation of OR, RR, RD with the baseline risk are −0.70 (95% CI, −0.88 to −0.32), 0.05 (95% CI, −0.40 to 0.48), and −0.35 (95% CI, −0.69 to 0.13). Both the BGLMM and Spearman's $\rho$ suggest a strong dependence between OR and baseline risk, with the OR decreasing as baseline risk increases. There is little evidence of such dependence with the RR, and a modest decrease of the RD as baseline risk increases. Thus, in a real data example we find portability of the RR but not of the OR, showing how portability depends on the actual situation under study, and demonstrating the invalidity of Doi et al.'s generalization from oversimplified constant-OR models.

**3.4 What to expect in practice?**

All effect measures lack portability in some scenarios; our previous paper provided a scenario showing lack of OR portability[9]. In practice, one may still wonder whether ORs are less correlated with baseline risks than RRs. While we recognize that correlations are not ideal measures for examining these issues, we nonetheless hypothesized that the magnitude of the correlation between the measure and baseline risks is similar in most cases. To evaluate our hypothesis, we summarized Spearman's $\rho$ between RRs and baseline risks ($\rho_{RR}$) among 20,198 meta-analyses (each has $n \geq 5$ studies with binary outcomes), following a similar procedure to obtain Spearman's $\rho$ between ORs and baseline risks ($\rho_{OR}$) in Xiao et al.[9] For studies with at least one zero count, we added 0.5 to all cells of the 2×2 table[48]. We compared $\rho_{RR}$ with $\rho_{OR}$ for each meta-analysis in a scatter plot. To avoid clutter and to investigate the impact of the number of studies in a meta-analysis, the scatter plot was stratified by the meta-analyses with $n < 20$ and $n \geq 20$ studies.

In the scatter plots, each point represents an individual meta-analysis, the x-axis is $\rho_{OR}$, and y-axis is $\rho_{RR}$ (Figure 2). Overall, ORs and RRs have similar correlation patterns over 20,198 meta-analyses. Since most points are distributed around the diagonal line, many meta-analyses have similar $\rho_{OR}$ and $\rho_{RR}$ values. Linear regression fits between $\rho_{RR}$ and $\rho_{OR}$ (the orange line) also confirm the similarity between $\rho_{OR}$ and $\rho_{RR}$. The majority of correlations were negative. 75% of meta-analyses with $5 \leq n < 20$ studies and 83% of meta-analyses with $n \geq 20$ studies had negative $\rho_{OR}$ and $\rho_{RR}$. While some meta-analyses suggest portability in the RR but not OR, a similar number of meta-analyses produce an opposite impression because the distribution of $(\rho_{OR}, \rho_{RR})$ for all meta-analysis points are almost symmetric about the diagonal line. Suppose we take 0.3 as the threshold for negligible correlation when using the absolute value of Spearman's $\rho$[49]. Then the proportions of meta-analyses with negligible $\rho_{OR}$ but non-negligible $\rho_{RR}$ are $6.1\%$ $(5 \leq n < 20)$ and $5.4\%$ $(n \geq 20)$, and the proportions of meta-analyses with negligible $\rho_{RR}$ but non-negligible $\rho_{OR}$ are $6.3\%$ $(5 \leq n < 20)$ and $11.7\%$ $(n \geq 20)$. The proportion of meta-analyses with negligible correlations is slightly higher for RR and baseline risks, and the difference is notable when the number of studies > 20. This also explains the above-diagonal linear regression fit when both correlation coefficients are negative, i.e., 0 > $\rho_{RR} > \rho_{OR}$ in Figure 2(a) and (b).

In summary, among 20,198 meta-analyses in Cochrane Reviews, we found many similarities between the correlations of ORs versus RRs and baseline risks. Those similarities include 1) the magnitude of correlation is similar between ORs and RRs, and 2) both ORs and RRs are negatively correlated with baseline risks in most cases. However, the proportion of meta-analyses with negligible correlations is slightly higher for the RR and baseline risks compared to the OR and baseline risks. We nonetheless expect that in most cases, neither measure will be safely portable. We thus advise showing how effect measures vary with their possible modifiers, especially baseline risk. A sensible choice for that task is a BGLMM; detailed explanations and the code to obtain effects conditioning on baseline risks is in Xiao et al.[9]

**4. Discussion**

We have emphasized that, contrary to the impression given by Doi et al., no effect measure is generally portable; the OR is no exception. After carefully reviewing their arguments, we found that their conclusion was guaranteed by an oversimplified statistical model which made an inappropriately strong assumption about constant ORs (rather than empirical evidence), and by mistakes in both their statistical and causal interpretations. Non-collapsibility is a mathematical feature of the OR that limits its use to situations in which risks are low for all patients; in those cases it will approximate the RR and render the present dispute moot.

Logistic regression does have important statistical advantages over log-binomial regression and can accommodate OR variation via product terms. But numerical niceties of the logistic regression model do not guarantee the OR's empirical portability. Using a real-world meta-analysis example, we showed that it is possible to have exactly the opposite of what Doi et al. claim: a setting in which the OR is not portable but the RR is portable. A large-scale comparison between the correlation with baseline risks of ORs and RRs further illustrates that the portability of both ORs and RRs vary across settings. Thus, ORs are *not* universally portable, and it is a mistake to replace RRs with ORs based on the alleged portability of the latter.

Choice of effect measure should consider interpretability and collapsibility as well as statistical ease. Doi et al.'s view that RRs are preferred over ORs "is a result more of tradition" is simply wrong: the preference for measures like RR and RD stem from their straightforward interpretations and greater relevance for clinical decision making. The OR offers only ease of statistical manipulation and estimation, which is likely why it is ardently promoted by those for whom such considerations override clinical or public health realism. But unlike RRs, ORs are difficult to interpret, non-collapsible, and can add bias if converted to RRs inappropriately. Furthermore, ORs can be more sensitive to sparse-data bias[50,51].

In conclusion, both ORs and RRs can be non-portable. As we emphasized in our previous paper[9], to address portability concerns one should report how effect measures vary with possible modifying factors such as baseline risks.

**Acknowledgement**

Research reported in this publication was partially supported by the National Center for Advancing Translational Sciences Award Number UL1-TR002494 (HC) and the National Library of Medicine Award Number R01LM012982 (MX and HC), R01LM013049 (RM) and R01LM012607 (YC) of the National Institutes of Health. The content is solely the responsibility of the authors and does not necessarily represent the official views of the National Institutes of Health.

Table 1: Generalized linear regression (GLM)* fits of odds ratio (OR) and risk ratios (RR) using all the data from Table 1 in Xiao et al.[1]

| | Data in Table 1B of Xiao et al. | | Data in Table 1A of Xiao et al.† | |
|---|---|---|---|---|
| Measure (model) | OR (logistic) | RR (log-binomial) | OR (logistic) | RR (log-binomial) |
| **Multiple regression model with interaction‡** | | | | |
| X | 2.67 | **2.00** | **2.67** | 2.00 |
| Z | 1.71 | **1.50** | **6.00** | 3.00 |
| X#Z | 1.31 | 1.00 | 1.00 | 0.67 |
| **Stratified regression models§** | | | | |
| X (Z=0) | 2.67 | 2.00 | 2.67 | 2.00 |
| X (Z=1) | 3.50 | **2.00** | **2.67** | 1.33 |
| Z (X=0) | 1.71 | 1.50 | 6.00 | 3.00 |
| Z (X=1) | 2.25 | **1.50** | **6.00** | 2.00 |
| **Univariate regression models (crude estimates)** | | | | |
| X | 3.00 | **2.00** | **2.25** | 1.50 |
| Z | 1.91 | **1.50** | **5.44** | 2.33 |

* GLM fits using the glm command in R with logit link (logistic model) or log link (log-binomial model); X, exposure; Z, stratum-specific variable; X#Z, interaction between exposure and stratum-specific variable; Y, outcome.

† The same data Doi et al. demonstrated in their response[2].

‡ Explanatory variables are X, Z and X#Z.

§ Explanatory variable is either a single X or single Z; Z=0, Z=1, X=0, X=1 correspond to the data subset scenario which Doi et al. referred as Z absent, Z present, X absent, X present.

Table 2: Meta-analysis estimates* of the association between X and Y if we assume the two strata Z=0 and Z=1 represent independent populations. OR and RR are homogeneous for different datasets.

| Association between X And Y | Data in Table 1B of Xiao et al. | | Data in Table 1A of Xiao et al.† | |
|---|---|---|---|---|
| | OR (95% CI) | RR (95% CI) | OR (95% CI) | RR (95% CI) |
| Z=0 | 2.67 (1.42, 5.02) | 2.00 (1.26, 3.17) | 2.67 (1.42, 5.02) | 2.00 (1.26, 3.17) |
| Z=1 | 3.50 (1.95, 6.29) | 2.00 (1.42, 2.81) | 2.67 (1.42, 5.02) | 1.33 (1.11, 1.61) |
| Meta-analysis summary | 3.09 (2.01, 4.74) | 2.00 (1.52, 2.63) | 2.67 (1.70, 4.17) | 1.54 (1.05, 2.06) |

* We used the rma command from the “metafor’’ package in R to fit random-effects meta-analysis models.

† The same data Doi et al. demonstrated in their response[2]. CI: compatibility intervals; OR: odds ratio; RR: risk ratio.

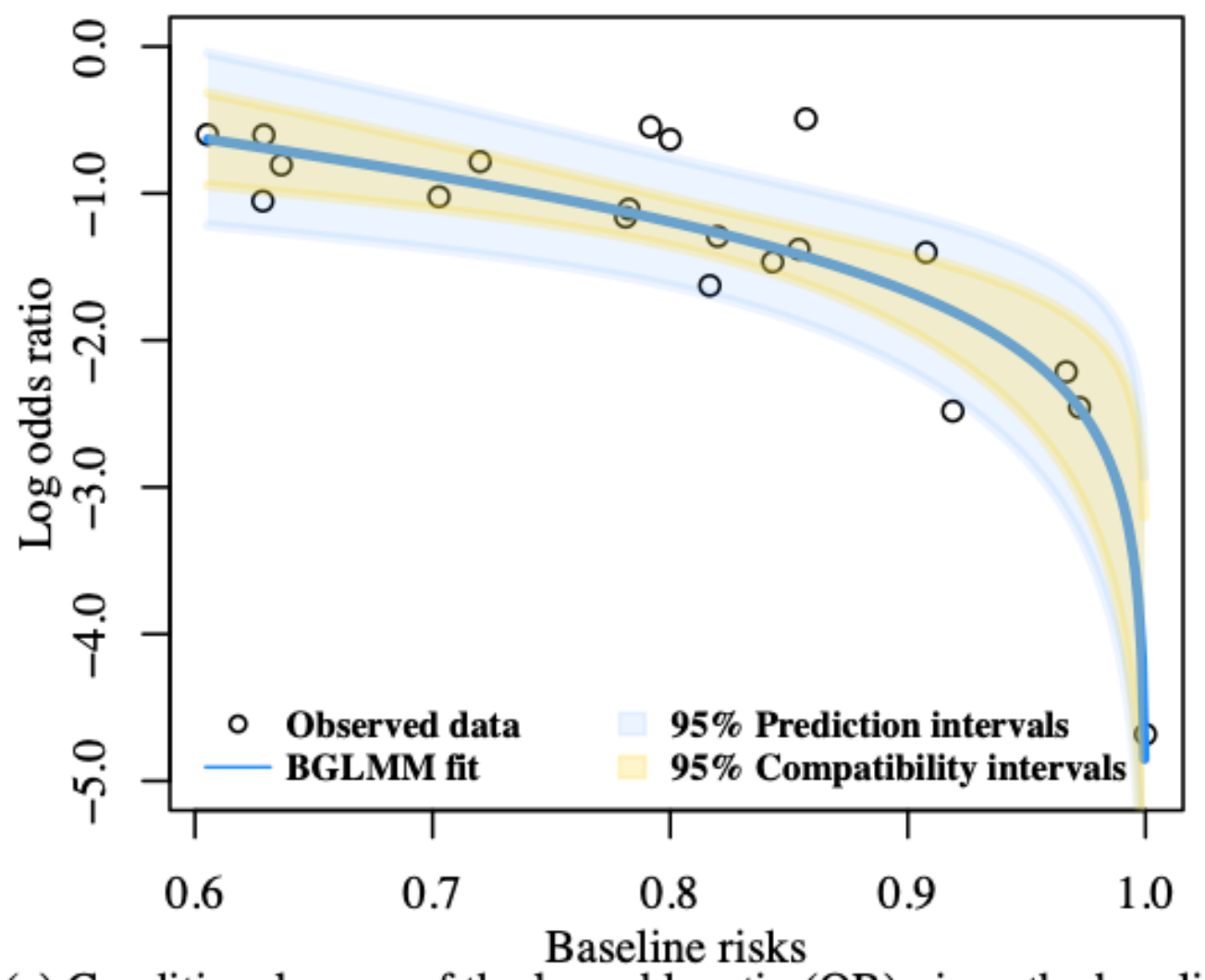

(a) Conditional mean of the log odds ratio (OR) given the baseline risks

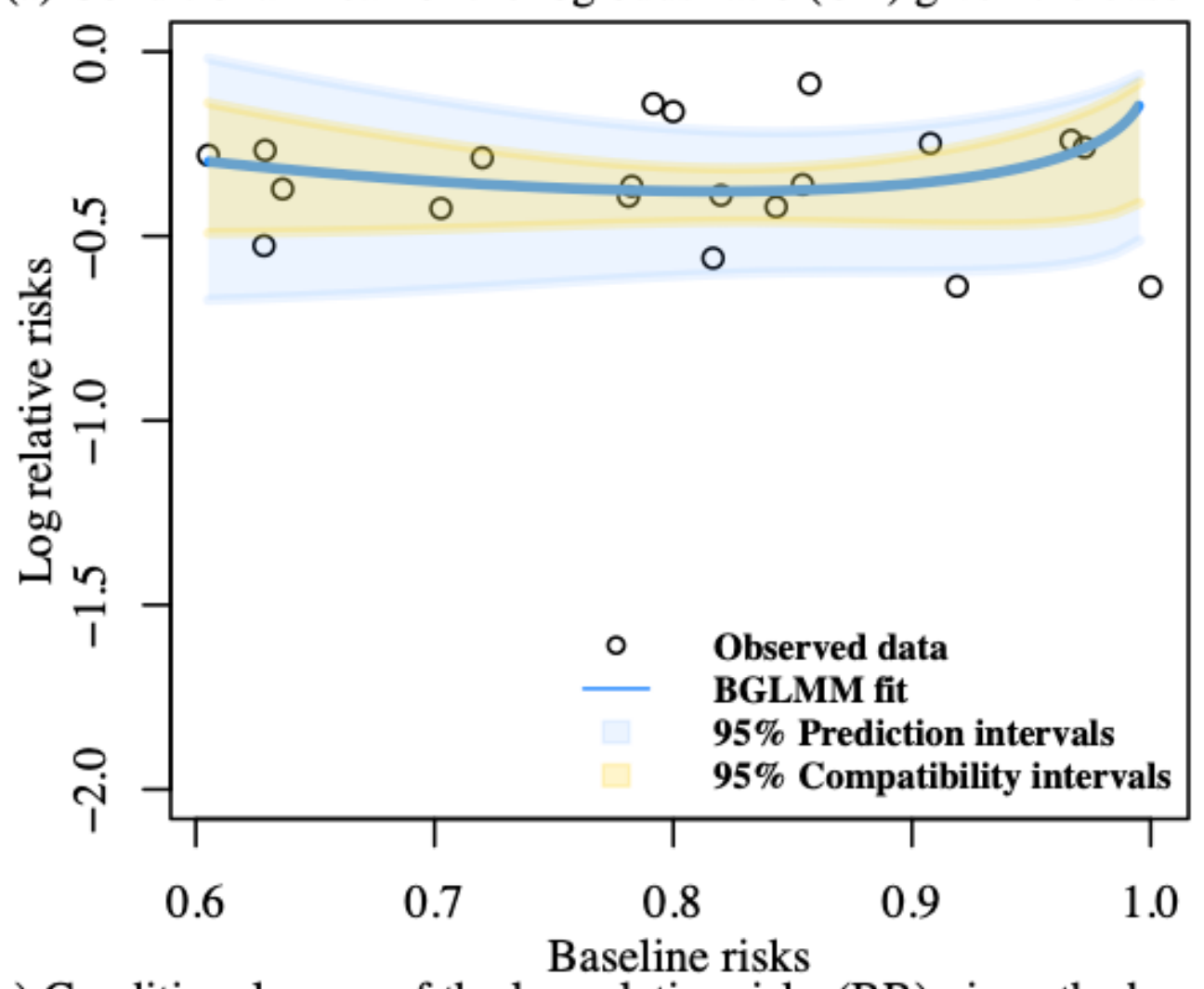

(b) Conditional mean of the log relative risks (RR) given the baseline risks

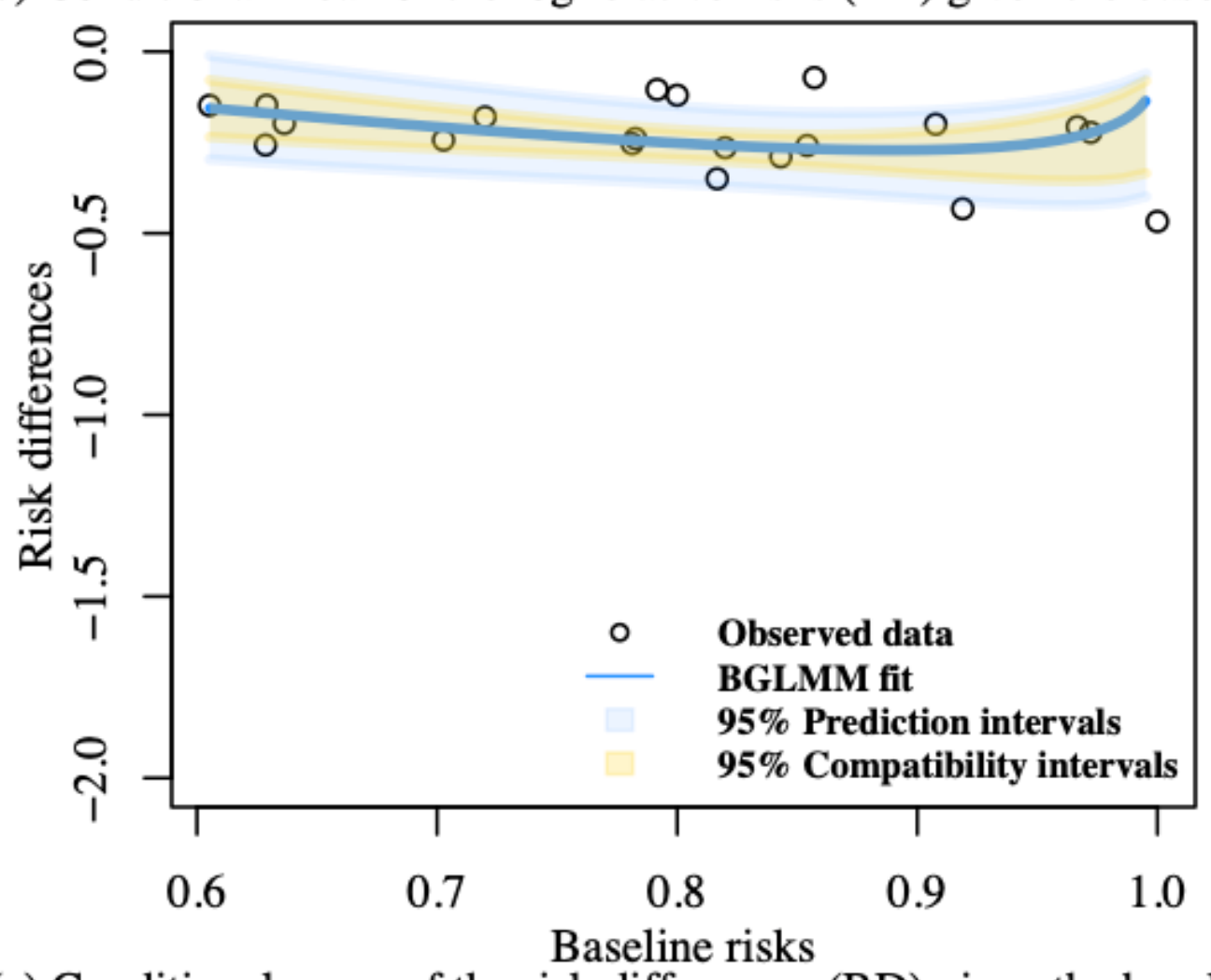

(c) Conditional mean of the risk differences (RD) given the baseline risks

Figure 1: Portability of three binary effect measures with increasing baseline risks in a real-world meta-analysis with 20 studies. BGLMM: bivariate generalized linear mixed-effects model.

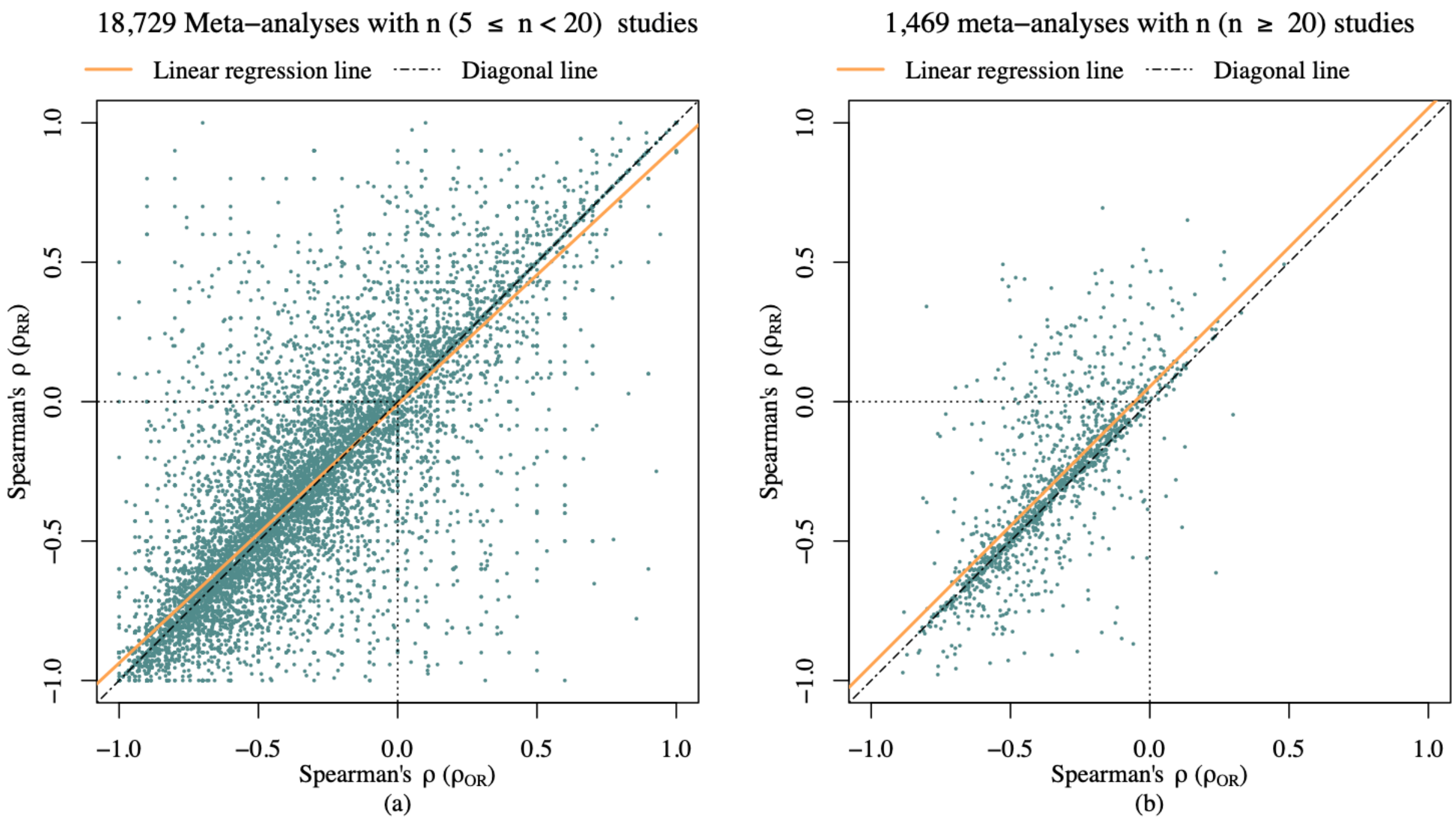

Figure 2: Scatter plot of Spearman's rank correlation coefficient (Spearman's $\rho$) between OR and baseline risk (horizontal axis) and Spearman's $\rho$ between RR and baseline risk (vertical axis) among 20,198 meta-analyses with ≥5 studies from Cochrane Reviews. OR: odds ratio; RR: risk ratio.

**Appendix – Code for Table 1 and Table 2**

**1. R code to reproduce Table 1.**

```
data.1B <- data.frame(X = rep(c(1,1,0,0),2), Y = rep(c(1,0,1,0),2), Z = c(rep(1,4),rep(0,4)))
data.1B <- data.1B[c(rep(1,60),rep(2,40),rep(3,30),rep(4,70),
        rep(5,40),rep(6,60),rep(7,20),rep(8,80)),]
data.1A <- data.frame(X = rep(c(1,1,0,0),2), Y = rep(c(1,0,1,0),2), Z = c(rep(1,4),rep(0,4)))
data.1A <- data.1A[c(rep(1,80),rep(2,20),rep(3,60),rep(4,40),
        rep(5,40),rep(6,60),rep(7,20),rep(8,80)),]
# function for computing results in table 1
tb1.result <- function(data,measure){
 link <- ifelse(measure=="OR","logit","log")
 m1 <- glm(Y~X+Z+X*Z,data=data,family = binomial(link = link))
 print(exp(coef(m1)[-1]))
 m11 <- glm(Y~X,data=data[data$Z==0,],family = binomial(link = link))
 print(exp(coef(m11)[-1]))
 m12 <- glm(Y~X,data=data[data$Z==1,],family = binomial(link = link))
 print(exp(coef(m12)[-1]))
 m13 <- glm(Y~Z,data=data[data$X==0,],family = binomial(link = link))
 print(exp(coef(m13)[-1]))
 m14 <- glm(Y~Z,data=data[data$X==1,],family = binomial(link = link))
 print(exp(coef(m14)[-1]))
 m1x <- glm(Y~X,data=data,family = binomial(link = link))
 print(exp(coef(m1x)[-1]))
 m1z <- glm(Y~Z,data=data,family = binomial(link = link))
}
# first column
tb1.result(data.1B,"OR")
# second column
tb1.result(data.1B,"RR")
# third column
tb1.result(data.1A,"OR")
# fourth column
tb1.result(data.1A,"RR")
```

**2. R code to reproduce Table 2.**

```
library(metafor)
tb2ma <- function(ma.obj){
## function to reproduce table 2.
## ma.obj is the meta-analysis object fitted by rma command in the “metaphor” package.
  print(apply(round(exp(cbind(ma.obj$yi,ma.obj$yi-1.96*sqrt(ma.obj$vi),ma.obj$yi+1.96*sqrt(ma.obj$vi))),2),2,rev))
  print("Meta-analysis summary:")
  print(round(exp(c(ma1$beta,ma1$ci.lb,ma1$ci.ub)),2))
}
# first column
ma1 <- rma(ai=c(80,40),bi=c(20,60),ci=c(60,20),di=c(40,80),measure = "OR")
tb2ma(ma1)
# second column
ma2 <- rma(ai=c(80,40),bi=c(20,60),ci=c(60,20),di=c(40,80),measure = "RR")
tb2ma(ma2)
# third column
ma3 <- rma(ai=c(60,40),bi=c(40,60),ci=c(30,20),di=c(70,80),measure = "OR")
tb2ma(ma3)
# fourth column
ma4 <- rma(ai=c(60,40),bi=c(40,60),ci=c(30,20),di=c(70,80),measure = "RR")
tb2ma(ma4)
```